\documentclass[usenatbib]{mn2e} 
\usepackage{epsfig,graphicx,amssymb,amsmath,natbib}
\begin{document} 
\title[LGS for ELT: Efficient SH WFS Design]{Laser Guide Stars for Extremely Large Telescopes: Efficient Shack-Hartmann Wavefront Sensor Design using Weighted center-of-gravity algorithm} 
\author[L. Schreiber et al.]{L.~Schreiber,$^1$ \thanks{E-mail: laura.schreiber2@unibo.it} I.~Foppiani,$^1$ C.~Robert,$^2$ E.~Diolaiti,$^3$ J.-M.~Conan,$^2$ M.~Lombini$^3$\\
$^1$Alma Mater Studiorum Universit\`a di Bologna, Dipartimento di Astronomia, via Ranzani 1, I-40127 Bologna (Italy)\\
$^2$ONERA, DOTA-CC, 29 av. de la div. Leclerc BP 72, 92322 Chatillon Cedex (France)\\
$^3$INAF-Osservatorio Astronomico di Bologna, via Ranzani 1, I-40127 Bologna (Italy)}

\date{} 
\maketitle 
\begin{abstract}
Over the last few years increasing consideration has been given to the study of Laser Guide Stars (LGS) for the measurement of the disturbance introduced by the atmosphere in optical and near-infrared astronomical observations from the ground. A possible method for the generation of a LGS is the excitation of the Sodium layer in the upper atmosphere at approximately 90 km of altitude. Since the Sodium layer is approximately 10 km thick, the artificial reference source looks elongated, especially when observed from the edge of a large aperture. The spot elongation strongly limits the performance of the most common wavefront sensors. The centroiding accuracy in a Shack-Hartmann wavefront sensor, for instance, decreases proportionally to the elongation (in a photon noise dominated regime). To compensate for this effect a straightforward solution is to increase the laser power, i.e. to increase the number of detected photons per subaperture. The scope of the work presented in this paper is twofold: an analysis of the performance of the Weighted Center of Gravity algorithm for centroiding with elongated spots and the determination of the required number of photons to achieve a certain average wavefront error over the telescope aperture.
\end{abstract}
\section{Introduction}
Adaptive Optics (AO) systems may provide diffraction-limited imaging with ground-based telescopes, measuring the wavefront aberrations due to the atmosphere by means of a suitably bright reference star and compensating for these aberrations by means of a deformable mirror. In classical AO systems, based on a single natural reference star, such a star has to be located close to the target object, i.e. within the isoplanatic patch, so that the probability of finding it is quite low. This probability defines the fraction of the sky where an efficient correction can be achieved (sky coverage), and for near infrared wavelengths it is few percent \citep{Beckers}. In this context the potential power of Laser Guide Star (LGS) AO, proposed by \citet{Foy}, is clear. The basic idea is to generate a bright guide star in the upper atmosphere by means of a laser. Two alternative physical phenomena may be exploited for such a purpose: Rayleigh scattering at low-medium altitude, up to $H \sim$ 10-20 km \citep{Fugate}, or excitation of the atoms in the Sodium layer located at approximately $H \sim$ 90 km, characterized by a thickness of $\Delta H \sim$ 10 km \citep{ThompsonGardner, Gardner, Happer}. Although low-altitude Rayleigh scattering can provide much more backscattered light, the higher altitude of the mesospheric Sodium layer reduces the cone effect \citep{Foy}, i.e. maximizes the atmospheric volume sampled by the artificial source. 

\begin{figure}
\begin{center}
\includegraphics[scale=0.4]{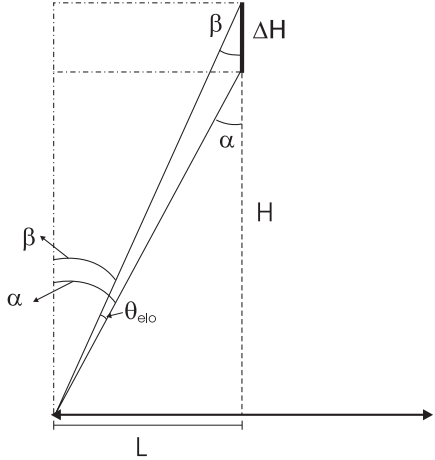}
\caption{Elongation of the LGS image at a distance $L$ (in this case at the edge of the telescope aperture) from the launch site. $\Delta H$ represents the thickness of the sodium layer, $H$ is the mean altitude of the layer and $\theta_{elo}$ is the angular elongation.}
\label{elongation}  
\end{center}
\end{figure}

The use of LGS in principle solves the problem of the lack of bright reference sources, but other difficulties arise. Apart from the above mentioned cone effect, that might be compensated using multiple reference sources \citep{ragazzoni1999}, the problem analyzed in this paper is related to the perspective elongation of the LGS image, due to the vertical extension of the artificial source. Referring, for instance, to a Shack-Hartmann (SH) WaveFront Sensor (WFS), the subapertures located far from the laser launcher see an elongated LGS image, pointing towards the laser launcher. The angular elongation is given by (see Figure \ref{elongation}) 
\begin{equation}\label{eq:elo1}
  \theta_{elo}=\alpha - \beta
\end{equation}
Since $\alpha$ and $\beta$ are given by 
\begin{align*}\label{eq:elo2}
\alpha=&\tan{\frac{L}{H}}& \beta=&\tan{\frac{L}{H+\Delta H}} 
\end{align*}
Equation \eqref{eq:elo1}, in the small angles approximation, becomes
\begin{equation}\label{eq:Na}
   {\theta_{elo}}\simeq \frac{L\Delta H}{H^2 + H\Delta H }  
\end{equation}
where $\Delta H$ is the thickness of the sodium layer, $H$ is the mean altitude of the sodium layer and $L$ is the distance between the pupil subaperture and the LGS launcher (Figure \ref{elongation}). This elongation can dramatically affect the capability to measure the wavefront, especially in the case of large diameter telescopes. 

On a 30-40 m class telescope, like the TMT \citep{TMT} and the E-ELT \citep{E-ELT}, the maximum elongation $\theta_{elo}$ varies between few arcsec and 10 arcsec, depending on the actual telescope diameter, on the Sodium layer properties and on the laser launcher position. 

Through numerical simulations, verified by means of analytical formulas, we compute the measurement error on the Optical Path Difference (OPD) per subaperture,   considering the elongation pattern on a SH WFS for the 42 m E-ELT \citep{E-ELT}, in the case of a laser guide star launched either from behind the secondary mirror (i.e. from the center of the pupil) or from the edge of the primary mirror. In Section \ref{sect:parameters} we describe the simulations and the input parameters. The performance analysis of a specific algorithm (Weighted Center of Gravity) is the subject of Section \ref{sect:subnoise}. In Sections \ref{sect:propagation} and \ref{sect:results2} we propagate the measurement error on the OPD per subaperture to RMS wavefront error over the whole pupil, hereafter referred to as WFE, assuming a modal least square wavefront reconstruction on a Karhunen-Loeve (K-L) basis. The resulting WFE is evaluated as a function of number of photons and Read-Out-Noise (RON). An estimation of the number of detected photons needed to achieve a pre-fixed WFE per LGS is given in Section \ref{sect:results2}.    


\section{Method} \label{sect:parameters}
\subsection{Sodium density profile}\label{sodium}
\begin{figure}
\begin{center}
\includegraphics[scale=0.5]{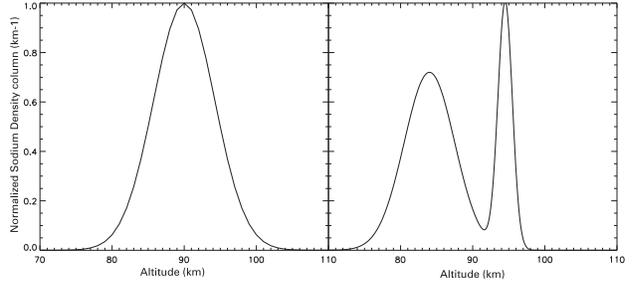}
\caption{The two density profiles of the Sodium layer considered in this paper.}
\label{profile}  
\end{center}
\end{figure}

In order to understand how the image of a LGS looks like as seen from each subaperture of a Shack-Hartmann wavefront sensor, it is important to choose a model of the Sodium density profile. The density of the Sodium layer can change by a factor of two or more depending on the telescope site and the time of the year. Its mean altitude can change on a time scale of hours \citep{sullivan}. We consider two different Sodium profile models (Figure \ref{profile}), with a shape similar to those presented in \citet{drummond}: 
\begin{itemize}
	\item Single Gaussian Sodium density profile, peak at $H = 90$ km and FWHM $\Delta H = 10$ km;
	\item Bi-modal Sodium density profile given by the sum of two Gaussians with the following characteristics: peak at $H = 84$ km and FWHM $\Delta H = 8.24$ km, peak at $H = 94.5$ km and FWHM $\Delta H = 2.35$ km.
\end{itemize}

\subsection{Modeling elongated spots}\label{elo_spots}

\begin{figure}
\begin{center}
\includegraphics[scale=0.4]{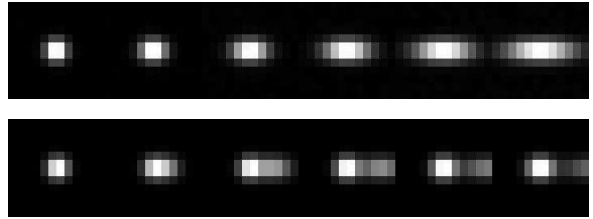}
\caption{Simulated LGS images with different elongations, as seen from subapertures at different distances from the laser projector, considering a Gaussian profile (top panel) and a bi-modal density profile (bottom panel).}
\label{spots}  
\end{center}
\end{figure}

The described Sodium profiles have been used to simulate the spots seen by a SH WFS. Some examples of simulated spots are shown in Figure \ref{spots}.

We consider two LGS launching schemes: from behind the secondary mirror (central projection) and from the edge of the primary mirror (lateral projection). Central projection gives a perspective elongation pattern with radial symmetry, while lateral projection leads to a non-symmetric pattern with maximum elongation two times larger than in the former option. The peak density of the Sodium layer is imaged at the center of the Field of View (FoV) of each subaperture. 

The LGS spot imaged by each subaperture is modeled as follows: 
\begin{itemize}
	\item the starting point is a line source of length calculated accordingly to Equation \ref{eq:Na} and with intensity distribution given by the adopted Sodium density profile;
	\item the line source is convolved by a seeing-like disk of angular size $\theta_{s} = 1.5$ arcsec, taking into account the round-trip propagation of the beam through the turbulent atmosphere (seeing = 1 arcsec). This angular size corresponds to the width in the non-elongated axis.
\end{itemize}
The subaperture diffraction ($\theta_{d} \sim 0.24$ arcsec) is not considered in the spot model, as it is negligible with respect to the seeing effect, that always dominates over diffraction since the AO correction at the laser wavelength is quite poor.

We generate sequences of instantaneous spots: each spot is characterized by a random jitter and is contaminated by photon noise and read-out noise (RON). The random offset is extracted from a Gaussian distribution, whose standard deviation may be tuned to simulate open or closed loop conditions. The photon noise is added to each pixel accordingly to a Poisson distribution with a mean value equal to the pixel intensity; the RON is extracted from a Gaussian distribution of zero mean.

Different values of focal plane sampling and subaperture FoV may be adopted in the generation of the sequence, in order to evaluate their impact on the WFS performance. 
Depending on the chosen subaperture FoV, the wings of the most elongated spots, especially for the the subapertures at the edge of the pupil, may fall outside the FoV of the subapertures themselves. In a real system, this would translate into a light contamination of adjacent subapertures, a drawback that may be overcome using simple light baffling methods, for instance a diaphragm on the LGS image in an intermediate focal plane before the WFS. This baffling is always implicitly assumed here, so that the limited subaperture FoV translates into a truncation of the spot wings, but not into a contamination of adjacent subapertures. The number of photons per subaperture referred to in the following includes also the photons blocked by the subaperture baffle.


\subsection{Algorithm performance evaluation} \label{parameters}

The sequence of instantaneous spots, typically formed by 500 realizations, is analyzed by a spot position measurement algorithm. The algorithm result is compared to the known true position, thus obtaining a sequence of position differences, the RMS of which is the figure of merit for the algorithm performance evaluation. Each instantaneous spot is tested for detectability. For this purpose two checks are implemented \citep{thomas}: the maximum intensity value of the spot is requested to be at least twice the RON and the calculated position has to be within $1-\sigma$ from the true position, where $\sigma$ is the standard deviation of the best fit Gaussian to the long-exposure spot, obtained by summing all the images in the sequence. When more than a certain fraction of images (in our simulations 50\%, following Thomas et al. 2006) are rejected, we consider the centroid measurements to have failed and not be reliable for those light conditions. 

\begin{table}
\begin{center}
\begin{tabular}{l l} 
\hline
Telescope parameters & D = 42 m\\
Relative obstruction & 0.3 linear \\
Non-elongated axis FWHM & $\sim$ 1.5 arcsec\\ 
Single Gaussian Profile & 10 km FWHM @90 km\\
Double Gaussian Profile & 8.24 km FWHM @84 km\\
                        & 2.35 km FWHM @94.5 km \\
No. of subapertures across D & 84\\
Subaperture FoV         & 12$\times$12 pixel\\
Pixel scale             & 0.75 arcsec/pixel\\
No. of photons per subap. & up to 2000\\
RON                     & 1, 3, 5, 7, 9, 11 $e^{-}$/pixel\\
Number of K-L modes     & 5600\\
\hline
\end{tabular}
\caption{Simulation parameters.}
\label{tab:parameters1}
\end{center}
\end{table}

The main simulation parameters are listed in Table \ref{tab:parameters1}. The values of some parameters, like the number of pixels per subaperture and the pixel scale, are the result of an optimization presented in Section \ref{results}. 
The performance of the algorithm has been evaluated considering subapertures at difference distances from the laser launcher, in order to have a complete map across the whole telescope pupil. 


\section{The Weighted Center of Gravity algorithm} \label{sect:subnoise}

\subsection{Algorithm description} \label{algo}

The estimation of the position of the spot on the focal plane of each subaperture of the SH WFS is generally performed by computing the Center of Gravity (CoG) of the spot. In this paper we analyze the performance of a slightly different algorithm, the Weighted Center of Gravity (WCoG), proposed by \citet{fusco}. This algorithm is a maximum likelihood estimator of the spot position in presence of Gaussian noise. The expression of the $x$ component of the spot position in a cartesian $(x,y)$ reference frame is given by
\begin{equation}\label{eq:fusco}
 C_{x} = \frac{\sum_{i,j}x_{i,j}W_{i,j}I_{i,j}}{\sum_{i,j}W_{i,j}I_{i,j}}
\end{equation}
where $x_{i,j}$ is the $x$ coordinate of pixel $(i,j)$, $I_{i,j}$ is the intensity of pixel $(i,j)$ and $W_{i,j}$ is a suitable weighting function. The $y$ component of the spot position is similar. The theoretical weighting function \citep{fusco} is the instantaneous noiseless spot image itself. In the case of a small spot jitter in the subaperture, the weighting function can be assumed equal to the mean spot, i.e. to the average of the 500 realizations in our case. 

\begin{figure}
\begin{center}
\includegraphics[scale=0.55]{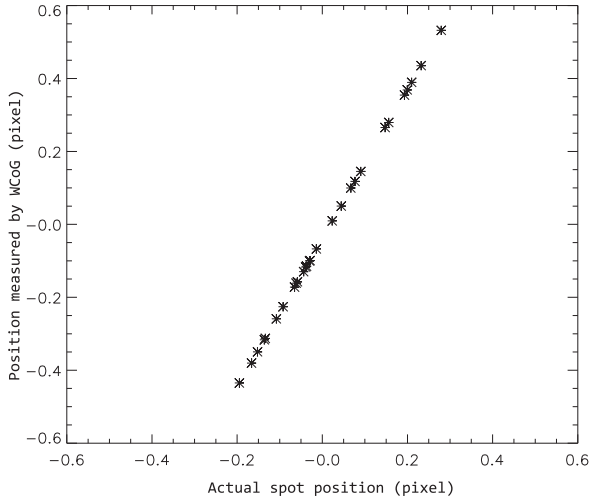}
\caption{Example of calibration curve for the WCoG algorithm. A template is moved in a set of known positions and then the spot position is measured by mean of the WCoG. The estimated positions are plotted against the actual ones. The resulting plot is fitted by a polynomial function, that is used to correct the WCoG measurement for any position inside the subaperture FoV.}
\label{linearita}
\end{center}
\end{figure}

The weighting function allows the attenuation of the noise effects, especially in the pixels with a low signal, but it also introduces a distortion on the position estimation, proportional to the distance of the actual spot position from the center of the weighting function. In \citet{thomas}, a correction factor, called $\gamma$ factor, has been analytically calculated in the case of a Gaussian spot. Since we are interested also in bi-modal and more general spot shapes and since a good knowledge of the instantaneous spot and weighting function are crucial to calculate the $\gamma$ factor \citep{sandrine}, instead of calculating a multiplicative $\gamma$ factor, we empirically derived a calibration curve to compensate for the biasing effects due to the weighting function. For a given spot template, depending on the Sodium profile, the result of the WCoG applied to the template not contaminated with noise is plotted against the actual spot position for several known positions inside the subaperture FoV. The resulting plot is the calibration curve that allows the compensation of the bias. The spot template can be thought of as the average spot over a period of time over which the Sodium profile is approximately constant. The calibration curve depends on the spot elongation and therefore changes with the subaperture position in the pupil. For each subaperture we find a curve like the one shown in Figure \ref{linearita}. The calibration curve is fitted by a second order polynomial, which is used to retrieve the correct spot position for any position inside the FoV. This method easily allows the compensation of non-linear effects, due for instance to the truncation of the spot at the subaperture FoV edge or to sampling effects like in a quad-cell scheme, as discussed in a paper in preparation \citep{centroidi}.

\begin{figure}
\begin{center}
\includegraphics[scale=0.55]{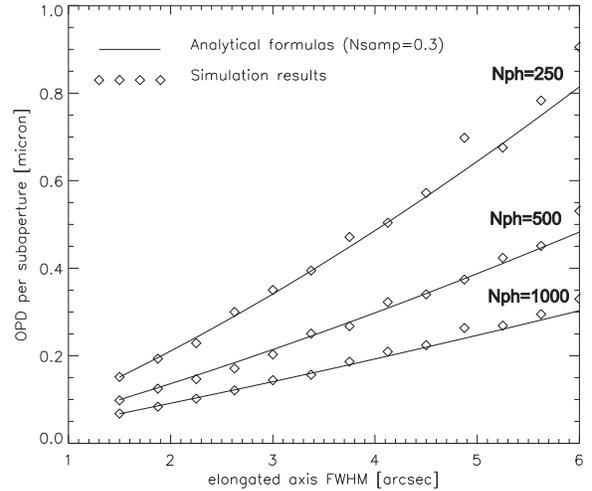}
\caption{Comparison between OPD measurement error as a function of the elongation value, parametric in the number of photons, calculated by numerical simulations (symbols) and by the analytical formula (continuous line). Central laser projection, RON = 3 $e^{-}$/pixel.}
\label{analytical}  
\end{center}
\end{figure}

The results of the numerical simulations have been verified using the analytical formulas of \cite{sandrine}; the parameter $N_{samp}$ of that paper, defined as the number of pixels per FWHM of the subaperture diffraction limited PSF, is assigned here the value $N_{samp}\sim 0.3$ (Figure \ref{analytical}).

A good agreement has been found between the simulations and the analytical formulas. A slight disagreement is found in the case of low number of photons, probably because of the increased measurement noise due to the presence of negative pixels. 


\subsection{Algorithm optimization and performance} \label{results}

The measurement error on the OPD per subaperture is taken as a figure of merit for the optimization of the algorithm parameters and for the performance evaluation. 
The basic parameters are the focal plane sampling and the FoV of the subaperture. Given a subaperture FoV, the calculation may be restricted to a smaller window, to optimize the performance on particular Sodium profiles, like the bi-modal, as will be discussed later in this Section. For the optimization of the parameters, a reasonably high signal-to-noise ratio case has been considered, corresponding to $N_{ph} = $500 and RON = 3 $e^{-}$/pixel. A central laser projection scheme has been assumed.

\begin{figure}
\begin{center}
\includegraphics[scale=0.55]{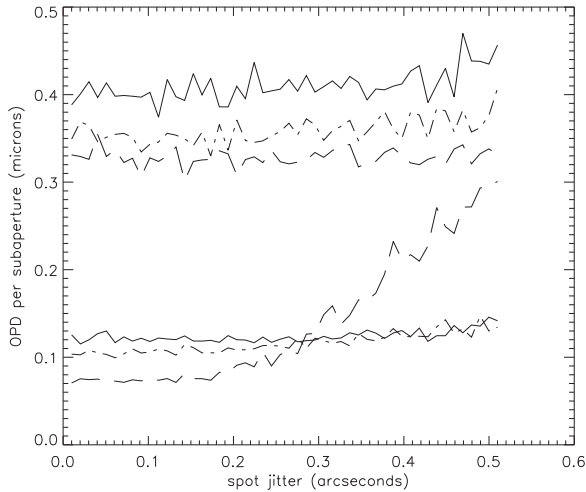}
\caption{OPD measurement error in an edge subaperture for different values of spot jitter and for a Gaussian Sodium profile. Three different pixel scales are compared: 0.75 arcsec/pixel (solid lines), 1 arcsec/pixel (dot-dashed lines) and 1.5 arcsec/pixel (dashed lines). The three upper curves refer to the elongated axis, the three lower to the non elongated one.}
\label{pixel_scale}  
\end{center}
\end{figure}

The first step of the optimization concerns itself with the sampling. Assuming a wide FoV, wide enough not to introduce any relevant truncation effect, the performance for three cases of sampling has been evaluted: 0.75, 1 and 1.5 arcsec/pixel. Considering the non elongated spot width (FWHM = 1.5 arcsec, as indicated in Table \ref{tab:parameters1}), the first case corresponds approximately to the Nyquist sampling of the spot itself. The performance has been evaluated in terms of error on the OPD measurement in an edge subaperture, as described in Section \ref{parameters}, for a Gaussian Sodium profile for different simulation sets, corresponding to different values of the spot jitter, ranging from closed to open loop conditions. Accordingly to \citet{Tyson}, the maximum spot jitter of 0.5 arcsec that we have considered a seeing condition of FWHM $\sim 1$ arcsec. The results are shown in Figure \ref{pixel_scale}, where the three upper curves refer to the elongated direction, while the three lower ones refer to the non elongated axis. From the elongated axis, one might infer that the coarser sampling is the best, because it improves the signal-to-noise ratio per pixel. However it is clear that the non elongated direction may be affected by undersampling effects for the coarser pixel scale, which increases the measurement error for the larger spot jitter cases. For this reason, in the following analysis the sampling of 0.75 arcsec/pixel has been adopted in the following, as it assures a constant error even in the worst seeing conditions. We do, however, just mention that in order to slightly relax the photon flux requirements without relevant drawbacks, a coarser sampling of 1 arsec/pixel also looks very attractive. 
 
\begin{figure}
\begin{center}
\includegraphics[scale=0.55]{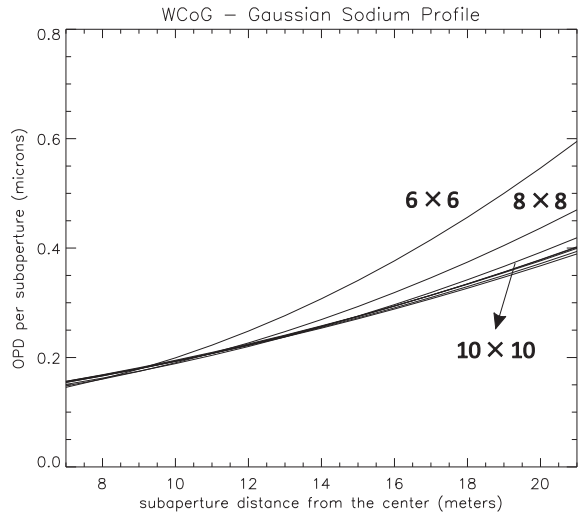}
\caption{OPD measurement error as a function of the subaperture distance from the center of the pupil (central laser projection) for different cases of subaperture FoV. Pixel scale 0.75 arcsec/pixel. Gaussian Sodium profile.}
\label{subap_dim_1}  
\end{center}
\end{figure}

\begin{figure}
\begin{center}
\includegraphics[scale=0.55]{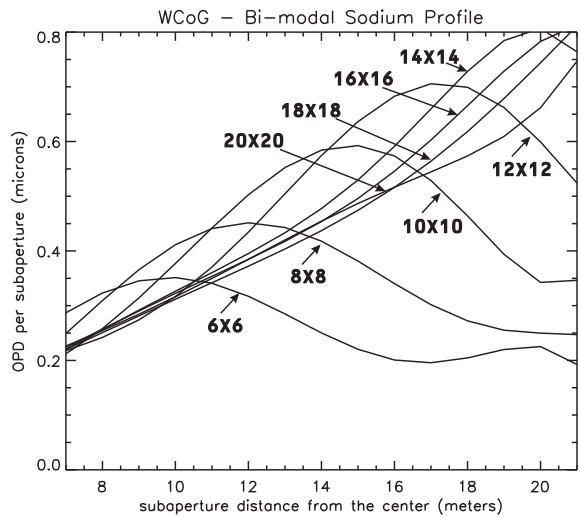}
\caption{Same as \ref{subap_dim_1}, but for a bi-modal Sodium profile. The FoV is centered on the sharper component of the spot.}
\label{subap_dim_2}  
\end{center}
\end{figure}

The second step of the optimization is concerned with the subaperture FoV. The analysis has been performed on the two reference Sodium profiles (single Gaussian and bi-modal), assuming the sampling previously derived (0.75 arcsec/pixel). The OPD measurement error has been evaluated as a function of the subaperture distance from the laser projector, considering different subaperture FoV values, ranging from $4 \times 4$ to $20 \times 20$ pixels. The results for the Gaussian Sodium profile and central laser projection are shown in Figure \ref{subap_dim_1}. The optimal FoV for this profile is of the order of $12 \times 12$ pixels, since wider FoVs do not further reduce the measurement error. For lateral laser projection, the optimal FoV is of the order of $24 \times 24$ pixels. The results for the bi-modal Sodium profile are shown in Figure \ref{subap_dim_2}. In this case a smaller FoV is clearly advantageous: this may be explained considering that a FoV as small as $6 \times 6$ pixels, centered on the sharper component of the bi-modal spot, translates into a mitigation of the spot elongation, although at the price of a loss of photon flux. This analysis shows that, for central laser projection, the optimal subaperture FoV, accounting for the different Sodium profiles considered here, is of the order of $12 \times 12$ pixels, but the calculation for the bi-modal case is more effective if restricted to a smaller window, extracted from the subaperture FoV. 

\begin{figure*}
\begin{minipage}{170mm}
\begin{tabular}{l l}
\psfig{figure=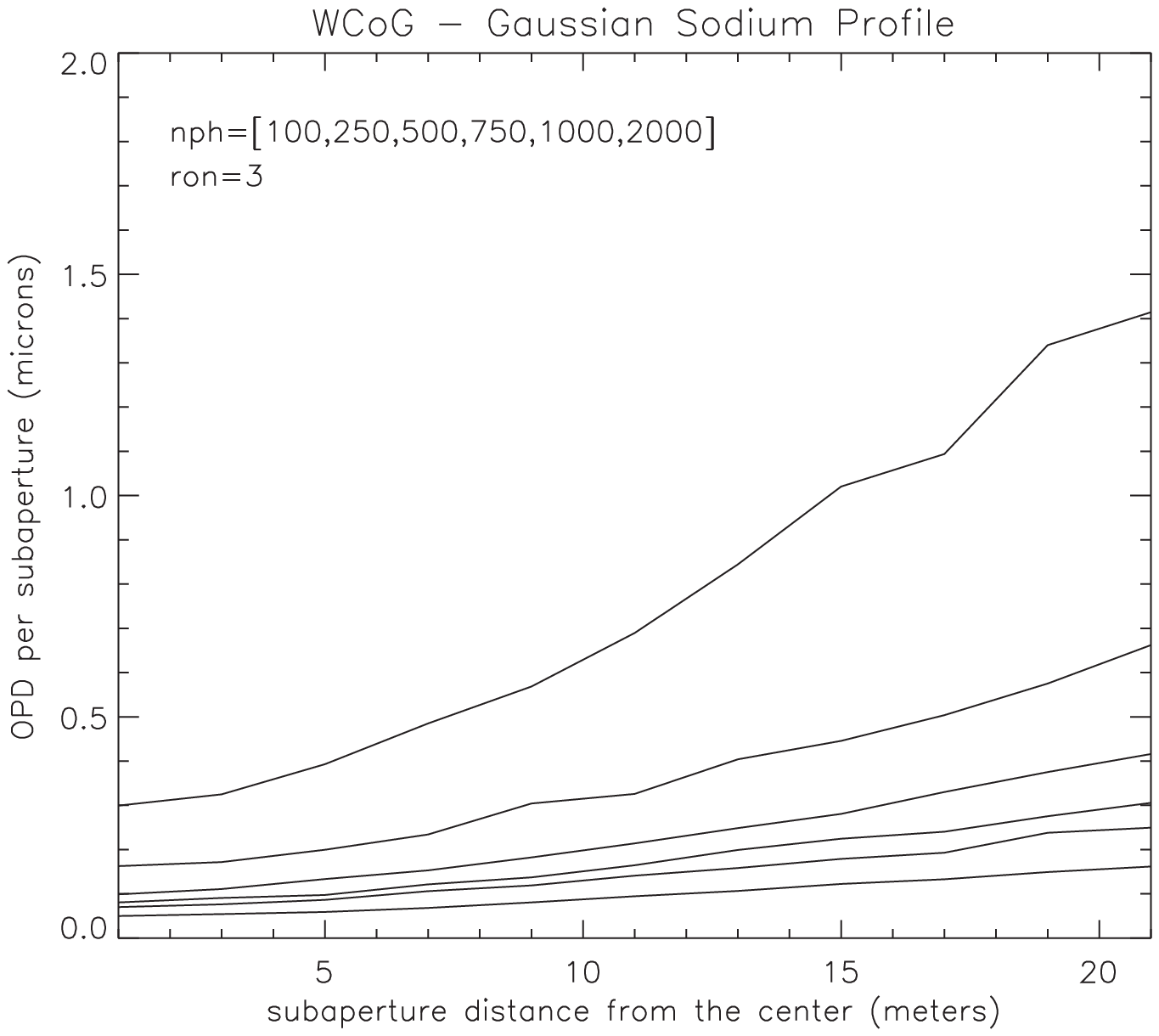,width=0.5\textwidth}&\psfig{figure=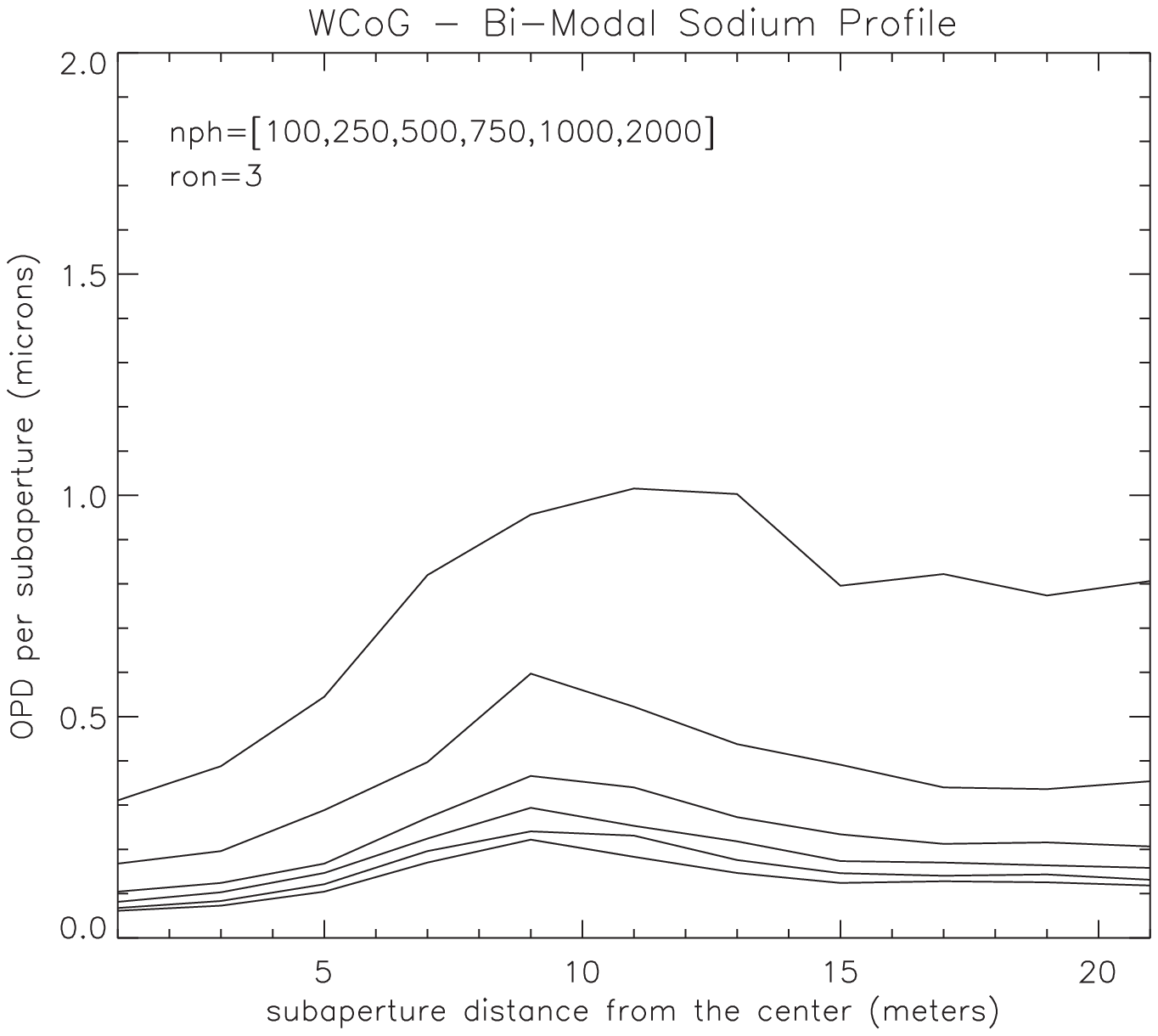,width=0.5\textwidth} \\      
\end{tabular}
\caption{OPD measurement error as a function of the distance of the subaperture from the pupil center, for central laser projection. The OPD error is parametric in terms of the number of photons per subaperture, ranging from $N_{ph} = $100 to $N_{ph} = $2000. The lowest curve in each plot refers to the highest number of photons. RON = 3 $e^{-}$/pixel, sampling 0.75 arcsec/pixel. Left: Gaussian Sodium profile, subaperture FoV = 12$\times$12 pixels. Right: bi-modal Sodium profile, window size 6$\times$6 pixel.}
\label{allnph}
\end{minipage}
\end{figure*}

\begin{figure*}
\begin{minipage}{170mm}
\begin{tabular}{l l}
\psfig{figure=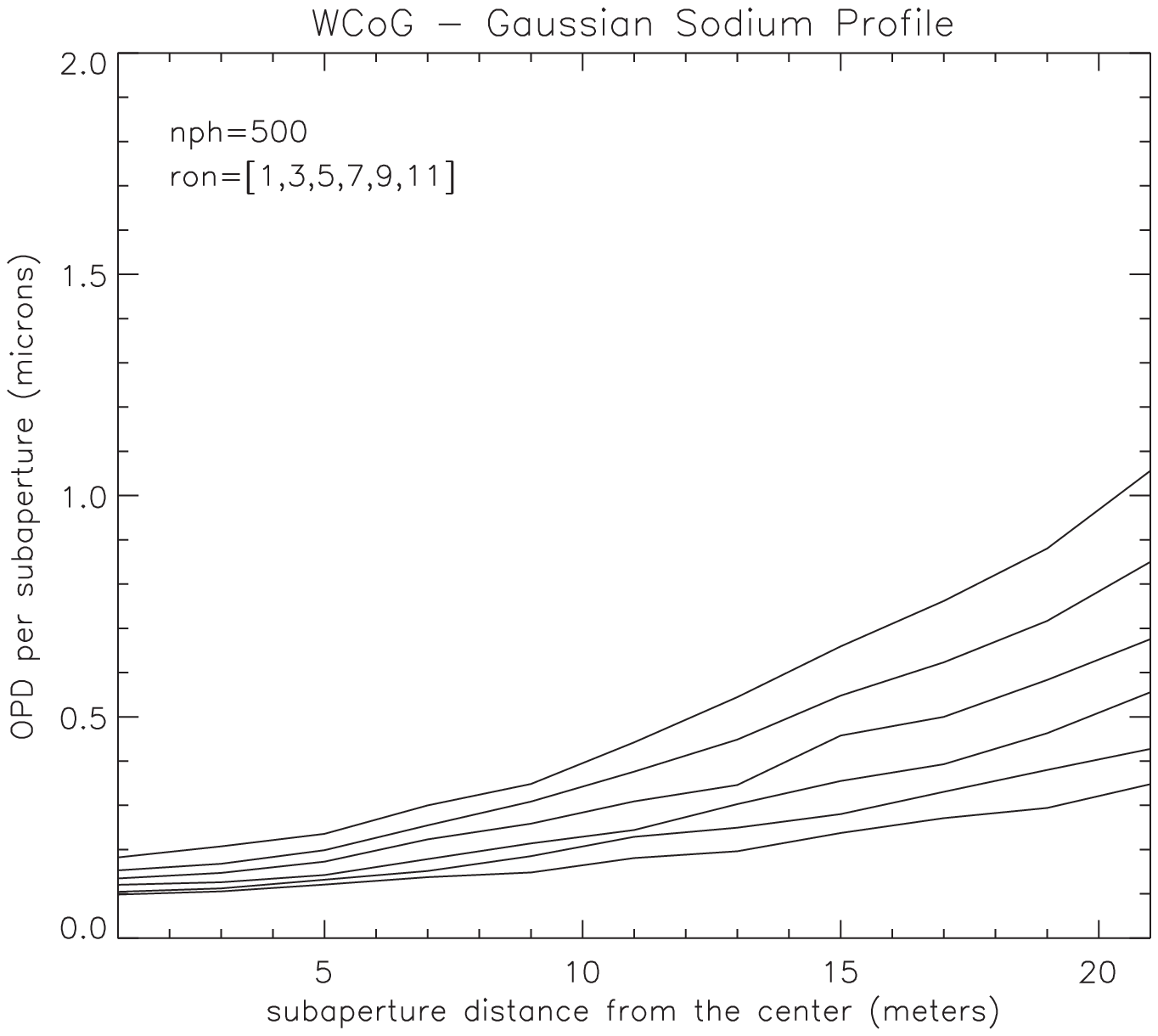,width=0.5\textwidth}&\psfig{figure=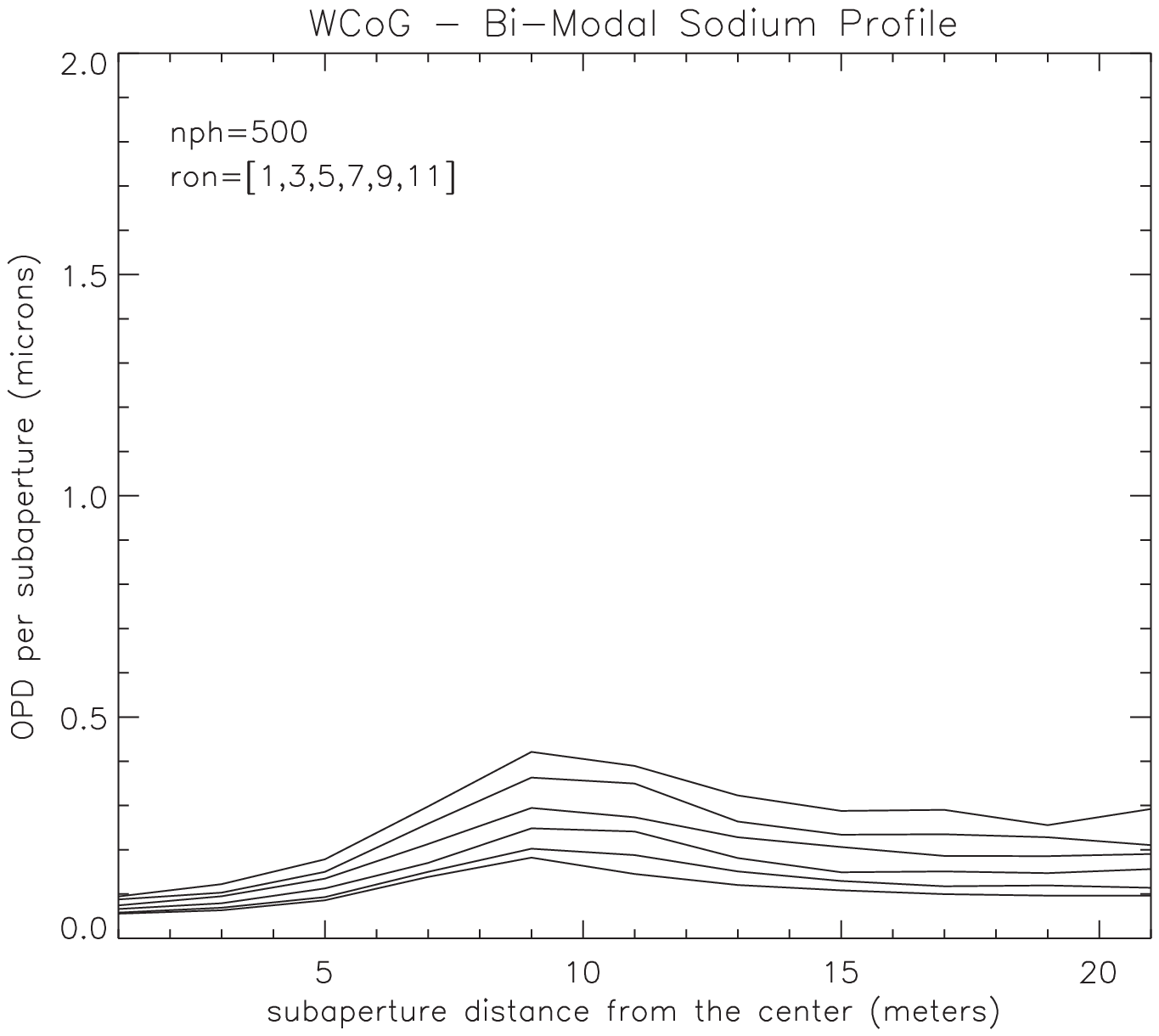,width=0.5\textwidth} \\      
\end{tabular}
\caption{Same as Figure \ref{allnph}, but curves are parametric in terms of RON, ranging from RON = 1 $e^{-}$/pixel to RON = 11 $e^{-}$/pixel. The lowest curve in each plot refers to the smallest value of RON.}
\label{allron}
\end{minipage}
\end{figure*}

The WCoG algorithm performance has been then evaluated for different values of photon flux and RON with the parameters optimized before (0.75 arcsec/pixel sampling, $12 \times 12$ pixels FoV with windowing down to $6 \times 6$ for bi-modal profile). The results are shown in Figures \ref{allnph} and \ref{allron}.

\section{Noise propagation} \label{sect:propagation}

The wavefront on the whole pupil is reconstructed by means of the OPD measurements per subaperture and hence the noise associated to the measurements is propagated to the reconstructed wavefront. The aim of this section is the translation of the OPD measurement errors calculated in Section \ref{sect:subnoise} into RMS wavefront error (WFE). This propagation is performed assuming that the wavefront may be expanded as a linear combination of K-L polynomials; the considered number of modes, $N=5600$, corresponds approximately to the total number of subapertures on the pupil of a telescope with the specifications listed in Table \ref{tab:parameters1}.

The OPD measurement problem and the wavefront reconstruction process can be expressed in a matrix-algebra framework \citep{Rousset}. The measurement equation is 

\begin{equation}\label{eq:np2}
   \mathbf{s} = \mathbf{Da}+\mathbf{n}
\end{equation}

where $\mathbf{a}$ is a vector gathering the $N$ unknown coefficients of the K-L modes, \textbf{s} contains a set of $2M$ OPD measurements ($M$ is the number of subapertures) and \textbf{n} is a vector of measurement errors. The matrix \textbf{D} is a $2M \times N$ interaction matrix giving the wavefront sensor response to the K-L modes; apart from a normalization constant, its $n$-th column contains the average value of the derivative of the $n$-th K-L mode for each subaperture in one direction. The matrix \textbf{D} is assumed to be sorted alternating the $x$ and $y$ derivatives, although any other ordering is acceptable, provided the vector \textbf{s} is ordered accordingly.

We consider a standard least-square reconstructor, meaning that the reconstruction matrix is simply the generalized inverse of \textbf{D}. Hence from Equation \eqref{eq:np2} we can write

\begin{equation}\label{eq:NP1}
  \mathbf{a_{est}} = \mathbf{Bs}
\end{equation}

where \textbf{B} is given by

\begin{equation}
 \mathbf{B}=\left(\mathbf{D^TD}\right)^{-1}\mathbf{D^T}.
\end{equation}

If the measurement noise \textbf{n} has a covariance matrix $\mathbf{C_s}$, the covariance matrix of the K-L coefficients is $\mathbf{B}\mathbf{C_{s}}\mathbf{B^{T}}$ and the wavefront error variance is the sum of the variances of the modal coefficients:

\begin{equation}
  \sigma^{2}_{a}=trace\left(\mathbf{B}\mathbf{C_{s}}\mathbf{B^{T}}\right).
  \label{eq:WFE}
\end{equation}

\begin{figure}
\resizebox{\hsize}{!}{\includegraphics{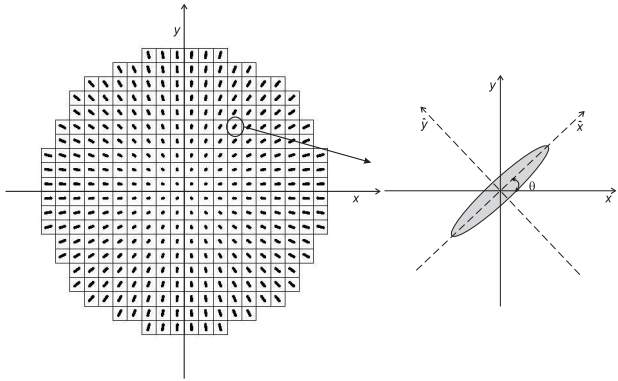}}
\caption{Elongation pattern for a laser guide star projected from behind the telescope secondary mirror. $(\hat{x},\hat{y})$ is a reference frame defined by the spot elongation direction and obtained by rotating the frame $(x, y)$ by an angle $\theta$.}
\label{fig:geomelo}  
\end{figure}

The noise covariance matrix $C_s$ is a block matrix of size $2M\times2M$ and can be easily constructed with the following prescriptions. The measurement noise in different subapertures is statistically uncorrelated, so the covariance matrix is a block diagonal matrix. Furthermore, if the measurement of the spot position could be performed with respect to the principal axes $(\hat{x}, \hat{y})$ of the elongated spot (Figure \ref{fig:geomelo}), then the measurement noise in the two axes would be uncorrelated. However the spot position measurement is performed with respect to the axes of the $(x, y)$ cartesian reference frame and the measurements errors in $x$ and $y$ are generally correlated: they may be retrieved from the measurement errors in the $(\hat{x}, \hat{y})$ reference frame applying a rotation matrix. The diagonal block referring to the $j$-th subaperture in the covariance matrix $C_s$ has the following structure:

\begin{equation}\label{eq:block}
\begin{pmatrix}
  \sigma^2_{\hat{x}_j}\cdot a^2_j + \sigma^2_{\hat{y}_j}\cdot b^2_j  &   \sigma^2_{\hat{x}_j}\cdot a_j \cdot b_j - \sigma^2_{\hat{y}_j}\cdot   a_j \cdot b_j\:\\&\\
  \:\sigma^2_{\hat{x}_j}\cdot a_j \cdot b_j - \sigma^2_{\hat{y}_j}\cdot   a_j \cdot b_j & \sigma^2_{\hat{x}_j}\cdot b_j^2 +     \sigma^2_{\hat{y}_j}\cdot a_j^2
  \end{pmatrix}
\end{equation}
where
\begin{flalign*}
a_j & =\cos\theta_j & &\\
b_j & =\sin\theta_j & & 
\end{flalign*}
and $\theta_j$ is the angle between the pre-defined global reference frame $(x, y)$ and the local reference frame $(\hat{x},\hat{y})$ defined in each subaperture by the direction of the elongation. An example of the elongation pattern on the whole aperture is shown in Figure \ref{fig:geomelo}. 

The calculation of the WFE, i.e. the square root of the wavefront error variance defined by Equation \ref{eq:WFE}, requires the evaluation of very large matrices with the adopted number of modes ($N = 5600$). For the design and optimization of the WFS, this evaluation has to be repeated for several cases of RON and of number of detected photons per subaperture. In order to speed up the process, we have applied a known property of the error propagation in a modal reconstruction approach \citep{Hardy}: the propagated WFE depends on the number of modes $N$ following a logarithmic law
\begin{equation}\label{eq:log}
 \text{WFE}\left(N\right) = a + b \log N
\end{equation}
where $a$ and $b$ are constants. The calculation of the WFE has been thus performed following Equation \ref{eq:WFE} for a relatively small number of modes, up to $N=465$, and the results have been fitted by the logarithmic curve described by Equation \ref{eq:log}; the best fit curve has been then extrapolated to the desired number of modes ($N=5600$), representing essentially the maximum number of independent modes measurable with the telescope specified in Table \ref{tab:parameters1}. This method has been verified on a system with just 20 subapertures across the diameter: an excellent match has been found between the full computation of the WFE following Equation \ref{eq:WFE} and the approximated result obtained by the extrapolation of the best fit logarithmic curve. 

\begin{figure}
\begin{center}
\includegraphics[scale=0.55]{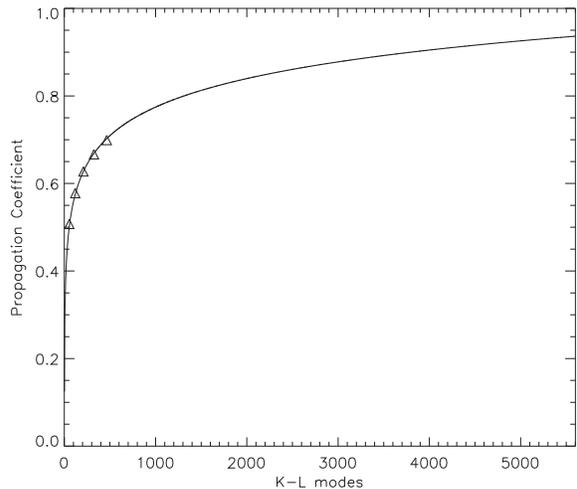}
\caption{Error propagation coefficient in case of no central obstruction as a function of the number of modes for a case without spot elongation (diagonal measurement noise covariance matrix). Triangles represent some of the results of the calculations up to $N=465$ modes; the extrapolated curve is the best fit logarithmic law described in the text.}
\label{fig:propcoeff}  
\end{center}
\end{figure}

The method proposed here has been validated also in a limiting, though relevant, situation: the case without spot elongation, where the OPD measurement error is the same in all subapertures. Here the measurement noise covariance matrix is diagonal, $\mathbf{C_{s}} = \sigma^{2}_{s} \textbf{I}$, and the propagated error is

\begin{equation}\label{eq:NP3}
 \sigma^{2}_{a}= trace\left(\mathbf{BB^T}\right)\sigma^2_s
\end{equation}

where $\sigma^2_s$ denotes the measurement noise variance and the term $trace(\mathbf{BB^T})$ is the so-called error propagation coefficient, usually of the order of or lower than 1. Figure \ref{fig:propcoeff} shows the behavior of the error propagation coefficient for a sample of mode numbers (up to $N=465$) and the best fit logarithmic curve (Equation \ref{eq:log}), extrapolated up to $N=5600$ modes. The error propagation coefficient estimated by this extrapolation is in very good agreement with the value found following the method proposed by \citet{rigaut}.

\section{Required number of detected photons} \label{sect:results2}

\begin{figure*}
\begin{minipage}{160mm}
\begin{tabular}{l l}
\psfig{figure=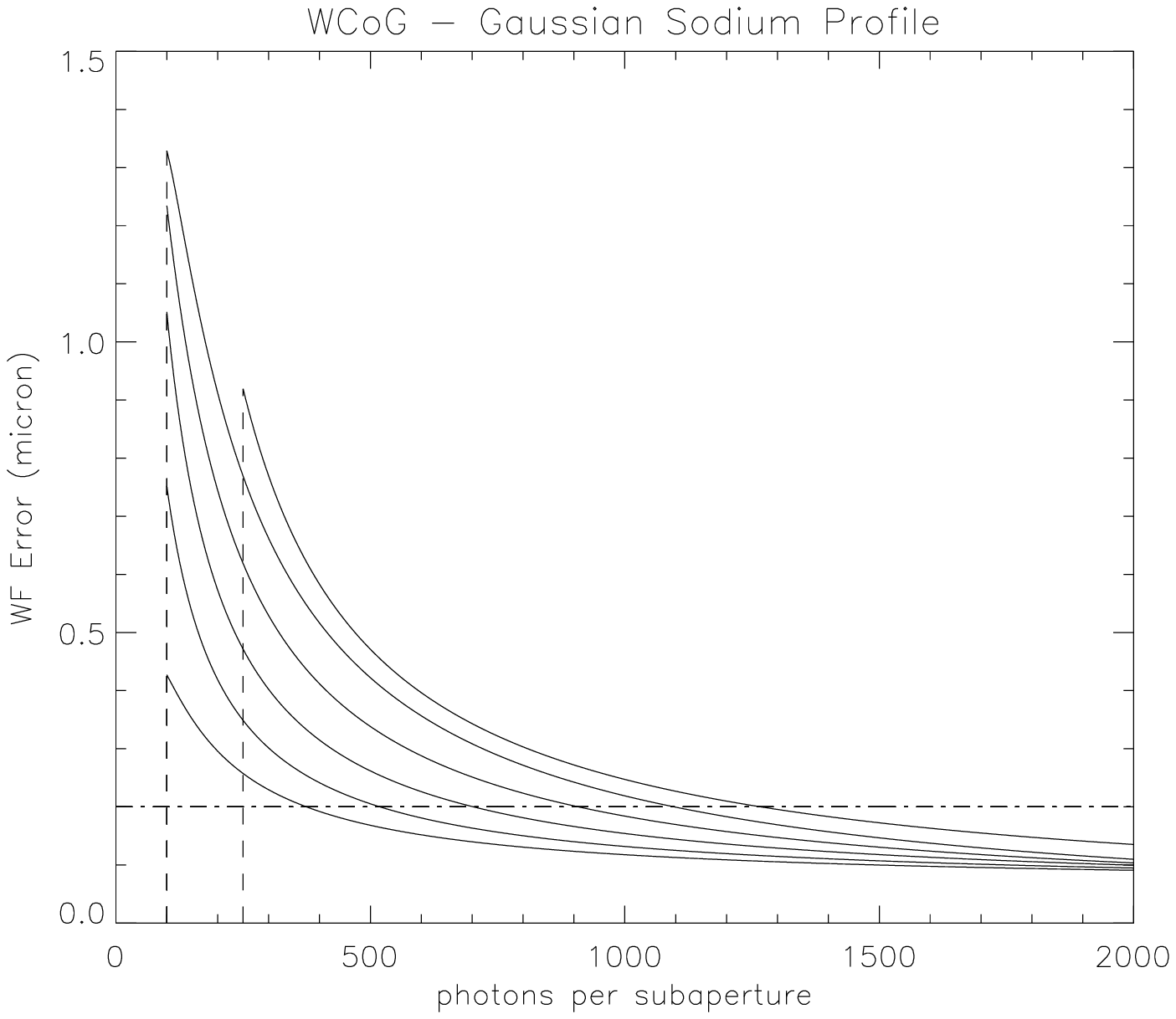,width=0.5\textwidth} &\psfig{figure=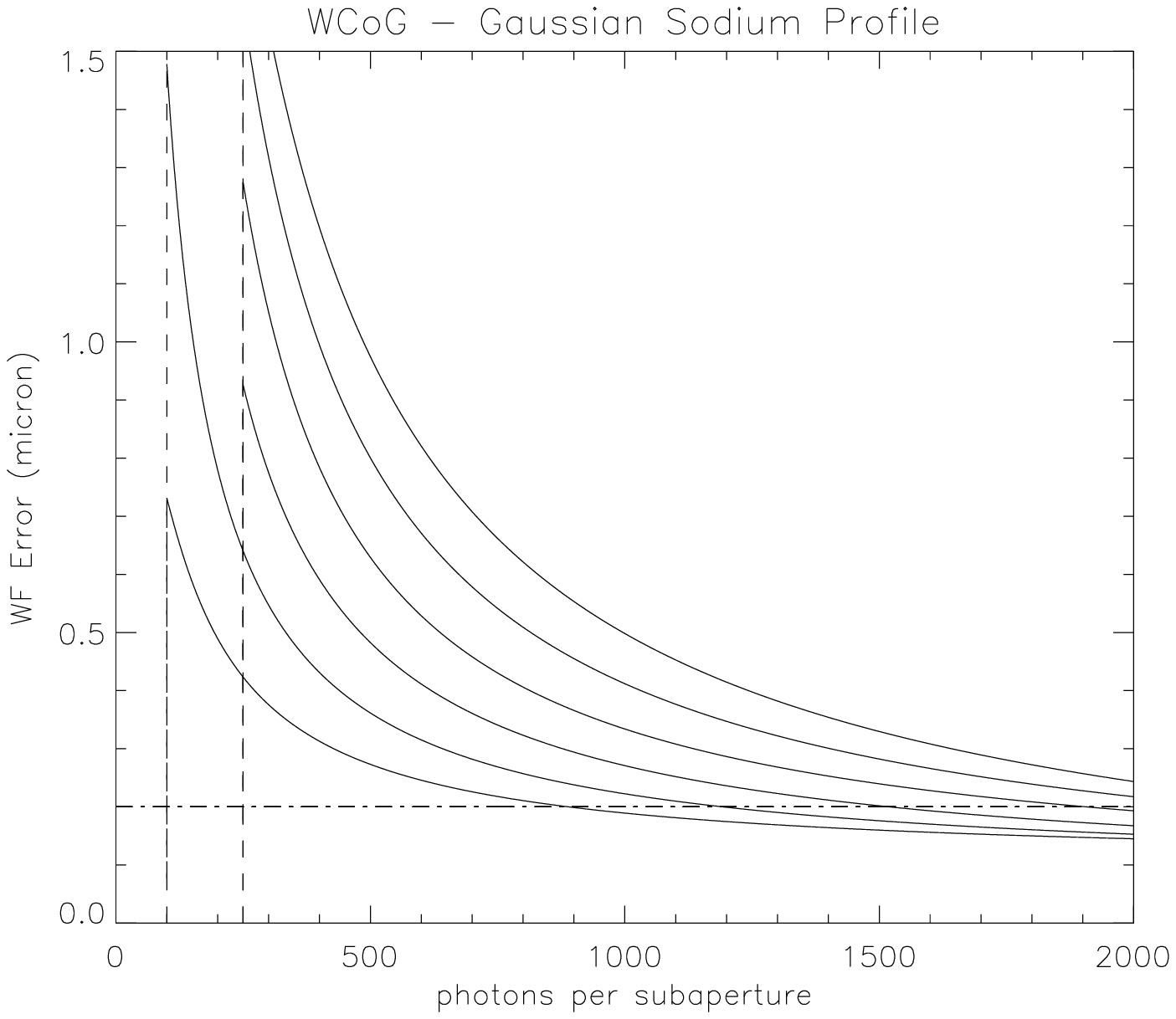,width=0.5\textwidth} \\      
\end{tabular}
\caption{Wavefront error vs number of detected photons per subaperture per frame for a Gaussian Sodium profile. The laser is projected from the center of the pupil (left panel) and from the edge (right panel). Different curves indicates different values of RON. The uppermost curve in each plot refers to the highest RON considered, i.e. RON = 11 $e^-/$pixel, and, in decreasing order, we find the curves relative to RON = 9, 7, 5, 3 and 1 $e^-/$pixel. The vertical lines represent the cut due to the detectability test described in Section \ref{parameters}.}
\label{gau}
\end{minipage}
\end{figure*}

\begin{figure}
\begin{center}
\includegraphics[scale=0.55]{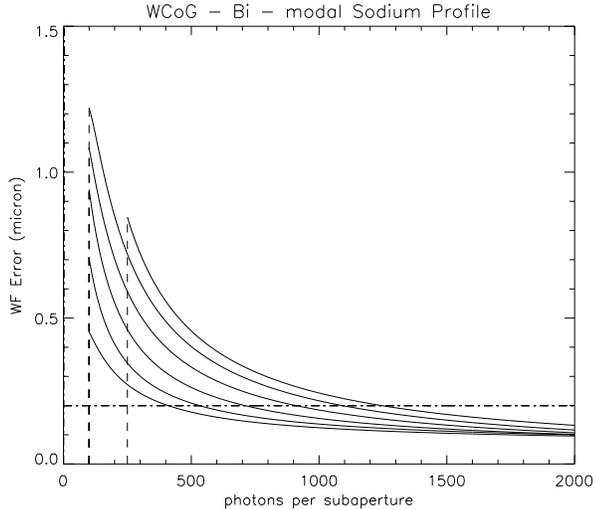}
\caption{Same as Figure \ref{gau} for a bi-modal Sodium profile and considering only central laser projection. A $6\times6$ pixels window has been applied to the bi-modal spot, in order to select only the sharper component.}
\label{bigau}  
\end{center}
\end{figure}

Applying the method explained in the previous Section, we calculate the WFE in a number of cases with different detected photons per subaperture per frame and different values of RON. Fitting the results for a given value of RON with a second order polynomial curve, we find the WFE as a continuous function of the number of detected photons. 
To evaluate the required number of detected photons per subaperture per frame, a target value for the WFE has to be defined. We based our study on the case of MAORY \citep{MAORY}, the Multi-Conjugate Adaptive Optics module for the European Extremely Large Telescope \citep{E-ELT}. In the error budget of MAORY, the target for the WFE in closed loop is 100 nm. Considering the number of deformable mirrors, the number of LGSs, the noise reduction due to the closed loop correction, the target open loop WFE applicable to the present single channel analysis is 200 nm. 

The analysis on the required number of photons has been performed in three relevant cases: 
\begin{itemize}
	\item  LGS projected from behind the telescope secondary mirror (maximum spot elongation $\theta_{elo}\sim 5$ arcsec) 
	\item  LGS projected from the edge of the primary mirror (maximum spot elongation $\theta_{elo}\sim 10$ arcsec)
	\item  No elongation, a condition achievable in principle by range gating \citep{Fugate} or dynamic refocus of a pulsed laser \citep{dynref}.
\end{itemize}

Figure \ref{gau} shows the results for the Gaussian profile of the Sodium layer and for the two laser projection configurations, central and lateral. The noise is clearly larger for lateral than for central projection, by a factor $\sim$ 2. The target WFE in each plot is shown by the horizontal dashed-dotted line. Figure \ref{bigau} shows the results for the bi-modal Sodium profile and central laser projection. In this case, the window used for the spot position calculations has been suitably optimized (see Section \ref{results}), in order to select the sharper component of the bi-modal spot. Comparing Figure \ref{gau} - left and \ref{bigau}, it may be noticed that the results are very similar, for a given laser projection configuration: for this reason, we consider in the following the Gaussian Sodium profile as the reference case. 

The results concerning the required number of detected photons to achieve the target value of WFE for different values of RON and in the three relevant cases mentioned above are gathered in Table \ref{tab:nph2}. A first evident conclusion is that the central laser projection is more effective than the lateral one from the point of view of photon requirements. However the latter may be advantageous for other reasons, in particular for the background contamination due to the Rayleigh scattering of the up-going laser beam in the lower atmosphere, that is stronger in the case of central projection. On the other hand, an additional aspect to investigate is the perspective elongation compared to the isoplanatic angle at the laser wavelength: the lateral launching is more sensitive to this problem than the central one. 
Finally it has to be noted that the considerable gain in the case without elongation is achievable by means of techniques such as dynamic refocus, which is technologically challenging, or range gating, that wastes a large fraction of the laser power.

\begin{table}
\begin{center}
\begin{tabular}{l c c c}
\hline
\textbf{RON} & \textbf{Central laser} & \textbf{Lateral laser} & \textbf{No elongation}\\
$\mathbf{1 e^-}$ &380&890&150 \\
$\mathbf{3 e^-}$ &520&1190&210 \\
$\mathbf{5 e^-}$ &700&1520&270 \\
$\mathbf{7 e^-}$ &910&1910&330 \\
$\mathbf{9 e^-}$ &1100&2200&410 \\
$\mathbf{11 e^-}$ &1270&2400&500 \\
\hline	
\end{tabular}
\caption{Number of detected photons per subaperture per frame required to achieve a WFE = 200nm per LGS. Three cases are considered: central laser projection, lateral projection and no spot elongation, a condition that might be achieved with a pulsed laser by range gating or dynamic refocusing. The results are presented for the Gaussian sodium profile. For the bi-modal sodium profile considered in this paper, assuming a proper windowing of the subaperture, the requested number of detected photons per subaperture is close to the results obtained in the Gaussian profile case.}
\label{tab:nph2}
\end{center}
\end{table}


\section{Conclusions and future work} \label{sect:Conclusions}

The effect of the LGS perspective elongation on the performance of a Shack-Hartmann WFS for a large aperture telescope has been analyzed in this paper. The Weighted Center of Gravity algorithm has been used for the spot position measurement in the subapertures. The OPD measurement error in the subapertures has been propagated to WFE (RMS wavefront error) across the whole pupil, assuming a modal least square wavefront reconstruction; a fast method to perform this error propagation has been proposed and validated, improving the efficiency of the calculation for a system with a very large number of subapertures. The method proposed here is based on a single channel approach, i.e. it allows the computation of the propagated wavefront error per single LGS, starting from the first order properties of the WFS (number of subapertures, number of detected photons per subaperture, RON, focal plane sampling, subaperture FoV, calculation window). 

The analysis presented in this paper has been tuned for the specific case of a MCAO module for the E-ELT, assuming a target WFE per LGS of 200 nm. Different Sodium profiles have been considered: Gaussian and bi-modal profiles. It has been found that, with a suitable optimization of the WCoG algorithm, performance is quite similar in the two cases. Two laser projection schemes have been considered: central and lateral projection. In this framework, assuming for instance a reasonable detector noise RON = 3 $e^-/$pixel, it has been found that approximately $N_{ph} = 520$ detected photons per subaperture per frame are required for the central laser projection. For comparison, the photon return of the Keck LGS system \citep{keck}, properly adapted to the case of the MCAO system for the E-ELT considering all relevant parameters, would correspond to approximately $N_{ph} \sim 70-180$; therefore the number of detected photons required to achieve our target WFE is a factor $\sim 3-7.5$ larger. For the lateral laser projection, the required number of photons is $N_{ph} = 1200$, i.e. a factor $\sim 2.2$ larger that for the central projection. The choice between central and lateral projection depends not only on the number of required photons, but also on other aspects (contamination due to the Rayleigh scattering of the laser light in the lower atmosphere, size of the isoplanatic angle compared to the perspective elongation), that are beyond the scope of this paper. A notable reduction in the required number of detected photons ($N_{ph} \simeq 210$) might be obtained by canceling the perspective elongation by means of a dynamic refocus scheme combined with a pulsed laser (Baranec \& Dekany 2008, Beckers et al. 2003, Georges et al. 2003), which is technologically challenging, or range gating \citep{martin}, that however wastes a non negligible fraction of laser power. Other approaches are under investigation to achieve similar results without fast moving parts \citep{tettine}. 

A detailed analysis of the influence of the RON has been performed. Assuming a more conservative value RON = 5 $e^-/$pixel, the required number of detected photons increases by a factor $\sim 1.4$ with respect to the case RON = 3 $e^-/$pixel. Another important aspect that has been analyzed is the number of pixels per subaperture, that determines the detector size. With typical Sodium profiles, assuming Nyquist sampling in the non elongated direction and central laser projection, the minimum number of pixels per subaperture is $12\times12$, so that 84 subapertures across the diameter (Table \ref{tab:parameters1}) corresponds to a detector size of $\sim 1000\times1000$ pixels, that doubles in the case of lateral laser projection. This detector format, especially combined with a RON = 3 $e^-/$pixel or better looks a demanding requirement. 

The analysis presented in this paper is based on the WCoG algorithm. However it is worth comparing this method to other algorithms (e.g. correlation maximization, quad-cell approach) with the aim of relaxing the requirements on the number of photons and on the detector specifications. Additional aspects to be considered in the evaluation of the algorithms are the variation of the Sodium profile with time and the impact of the differential aberrations between science and LGS channel, that usually depend on the Zenithal angle variation during the exposure. The Sodium profile variations imply continuous updating of the algorithm reference (the weighting function and the calibration curve in the case of the WCoG); the differential aberrations may translate into time dependent spot offsets in the subapertures, that need to be compensated or calibrated. The evaluation of different algorithms, considering these additional issues, is the subject of an additional work in preparation \citep{centroidi}.

 
\section*{Acknowledgements}
The authors thank R. Clare, N. Hubin, T. Fusco, M. Le Louarn, E. Marchetti, C. Petit, M. Tallon for useful discussions. 
This work was supported by the European Community (Framework Programme 6, ELT Design Study, contract N. 011863; Framework Programme 7, Preparing for the construction of the European Extremely Large Telescope, contract N. 211257).

\end{document}